\documentclass[aps,pra,reprint,twocolumn,superscriptaddress,showpacs,showkeys,floatfix,bibnotes]{revtex4-2}

\usepackage{graphicx}
\usepackage{epstopdf}
\usepackage{amssymb}
\usepackage{amsmath}
\usepackage{bm}
\usepackage{braket}
\usepackage{color}
\usepackage{makecell}
\usepackage{scalerel}

\usepackage[colorlinks,citecolor=blue]{hyperref}

\usepackage{float}

\begin{document}
	\title{Anomalous bulk-edge correspondence of nonlinear Rice-Mele model}
	\author{Chenxi Bai}
	\affiliation{Department of Physics, Zhejiang Normal University, Jinhua 321004, China}
	\author{Zhaoxin Liang}
	\thanks{zhxliang@zjnu.edu.cn}
	\affiliation{Department of Physics, Zhejiang Normal University, Jinhua 321004, China}
\begin{abstract}
Bulk-edge correspondence constitutes a fundamental concept within the domain of topological physics, elucidating the profound interplay between the topological invariants that characterize the bulk states and the emergent
edge states. A recent highlight along this research line consists of establishing bulk-edge correspondence under
the eigenvalue’s nonlinearity in a linear Hamiltonian by introducing auxiliary eigenvalues [\href{https://doi.org/10.1103/PhysRevLett.132.126601}{ T. Isobe {\it et al.,} Phys. Rev. Lett. 132, 126601 (2024)}]. The purpose of this work aims to extend Isobe’s analysis to uncover bulkedge correspondence of the eigenvalue’s nonlinearity in intrinsic nonlinear Hamiltonians. To achieve this, we
numerically solve the nonlinear Rice-Mele (RM) model and identify two distinct types of nonlinear eigenvalues:
the intrinsically nonlinear eigenvalues and the eigenvalue’s nonlinearity introduced through the incorporation of
auxiliary eigenvalues. Furthermore, we establish a form of bulk-edge correspondence based on these auxiliary
nonlinear eigenvalues, which we term the anomalous bulk-edge correspondence of a nonlinear physical system.
The concept of the anomalous bulk-edge correspondence defined herein provides an alternative perspective on
the intricate interplay between topology and nonlinearity in the context of bulk-edge correspondence.
\end{abstract}
	
\maketitle

\section{Introduction}
The bulk-edge correspondence stands as a cornerstone principle in topological physics~\cite{Xiao2010, Citro2023, Eschrig2011}, positing that the topological invariants of bulk states, such as the Chern number, determine the number and nature of edge states. There is currently a surge of interest in exploring the bulk-edge correspondence across a diverse spectrum of physical systems~\cite{Haldane2008, Raghu2008, Wang2008, Lu2014, Wu2015, Takahashi2018, Ozawa2019, Yasutomo2020, Yuto2022, Kariyado2015, Yang2015, Huber2016, Roman2016, Kawaguchi2017, Takahashi2019, Liu2020, Amin2018, Storkanova2021, Makino2022, Hu2022, Knebel2020, Yoshida2021, Mezil2017}, including interdisciplinary areas like meteorological systems \cite{Pierre2017}. The fascination with the bulk-edge correspondence stems from two primary drivers. First, within topological physics~\cite{Kane2005, 2Kane2005, Fu2007, Hasan2010, Qi2011, Ando2013, Andrei2006, 2Fu2007, Fu2006, Thouless1983, Schnyder2008,Zhu2020}, the bulk-edge correspondence is a hallmark feature of topological materials~\cite{Hatsugai1993, 2Hatsugai1993}, encompassing topological insulators, topological semimetals, and topological superconductors. It serves as a pivotal method for identifying and characterizing these materials. Second, the bulk-edge correspondence not only sheds light on the intrinsic connection between bulk and edge states, but also lays the theoretical groundwork for the application of topological materials. For example, in fields such as quantum computing~\cite{Kunst2018, Asboth2014, Graf2013} and spintronics~\cite{Hadad2018}, harnessing the edge states of topological materials facilitates low-loss and highly stable electron and spin transport. To date, the bulk-edge correspondence has predominantly been explored within the realm of linear quantum systems~\cite{Sadel2017, Yuce2018}.

Extending our understanding beyond linear quantum systems, the intricate interplay between topology and many-body effects~\cite{Jurgensen2021,Jurgensen2023,Viebahn2024,Cao2024,Cao2024a,Cao2024b} in quantum systems gives rise to a rich tapestry of topological many-body phenomena. This interplay has been the subject of intensive investigation in both fermionic~\cite{Citro2003, Requist2018, Nakagawa2018, Bertok2022, Stenzel2019} and bosonic systems~\cite{Berg2011, Qian2011, Grusdt2014, Zeng2016, Greschner2020}. These efforts are fueled by the rapid advancements in programmable materials and the increasing integration of topological principles into practical applications. A natural and compelling extension of this research is to delve into how the interplay between nonlinearity (or interactions more broadly) and topology influences the behavior of bulk-edge correspondence. This line of inquiry promises to uncover additional insights into the fundamental properties of bulk-edge correspondence and may pave the way for the discovery of novel topological states in these systems.

In the recent work~\cite{Isobe2024}, the authors have ventured into the realm of eigenvalue nonlinearity~\cite{Isobe2024, Cheng2024, Haldane1988, Yang2021, Yingxia2024, Zhang2021, Ronald2024} within the context of linear Hamiltonians. Their exploration offers a fresh perspective on the bulk-edge correspondence in the nonlinear quantum domain. A notable finding is that, when the nonlinearity is weak but not negligible, the topological edge states of the auxiliary eigenstates are inherited as physical edge states. This discovery underscores the deep connection between the auxiliary eigenvalues and physical systems. The work carries significant physical implications. It suggests that the method of introducing auxiliary eigenvalues can be generalized to systems belonging to different symmetry classes and dimensions, opening up new avenues for understanding and exploring the bulk-edge correspondence in a wider range of physical systems.

Despite the groundbreaking insights offered by the aforementioned study, its focus is restricted to eigenvalue’s nonlinearity within the framework of linear Hamiltonians. This naturally prompts the inquiry: What is the nature of the bulk-edge correspondence in the realm of nonlinear Hamiltonians~\cite{Li2023, Haldane1981} that exhibit nonlinear eigenvalues? This question stands as a fertile ground for exploration, as it holds the promise of uncovering novel mechanisms and phenomena within the topological landscape of nonlinear systems.

In this work, we shift our focus to the nonlinearity of the eigenvalues within nonlinear Rice-Mele (RM) Hamiltonians,
offering a distinct perspective compared to Ref.~\cite{Isobe2024}. Our investigation encompasses not only the eigenvalue spectrum, but also the bulk-edge correspondence. Utilizing numerical methods, we delve into how nonlinearity influences the stability of edge states, adopting an approach similar to Ref.~\cite{Obana2019}. To achieve this, we numerically solve the nonlinear RM model and identify two distinct types of nonlinear eigenvalues: the intrinsically nonlinear eigenvalues of nonlinear RM model and the auxiliary nonlinear eigenvalues introduced through the concept of auxiliary eigenvalues. Furthermore, we establish a form of bulk-edge correspondence based on these auxiliary nonlinear eigenvalues, which we term the ``anomalous bulk-edge correspondence” of a nonlinear physical system. The notion of ``anomalous” introduced here provides fresh insights into the interplay between topology and nonlinearity in the context of bulk-edge correspondence.

\begin{figure*}
\includegraphics[width=1.0\textwidth]{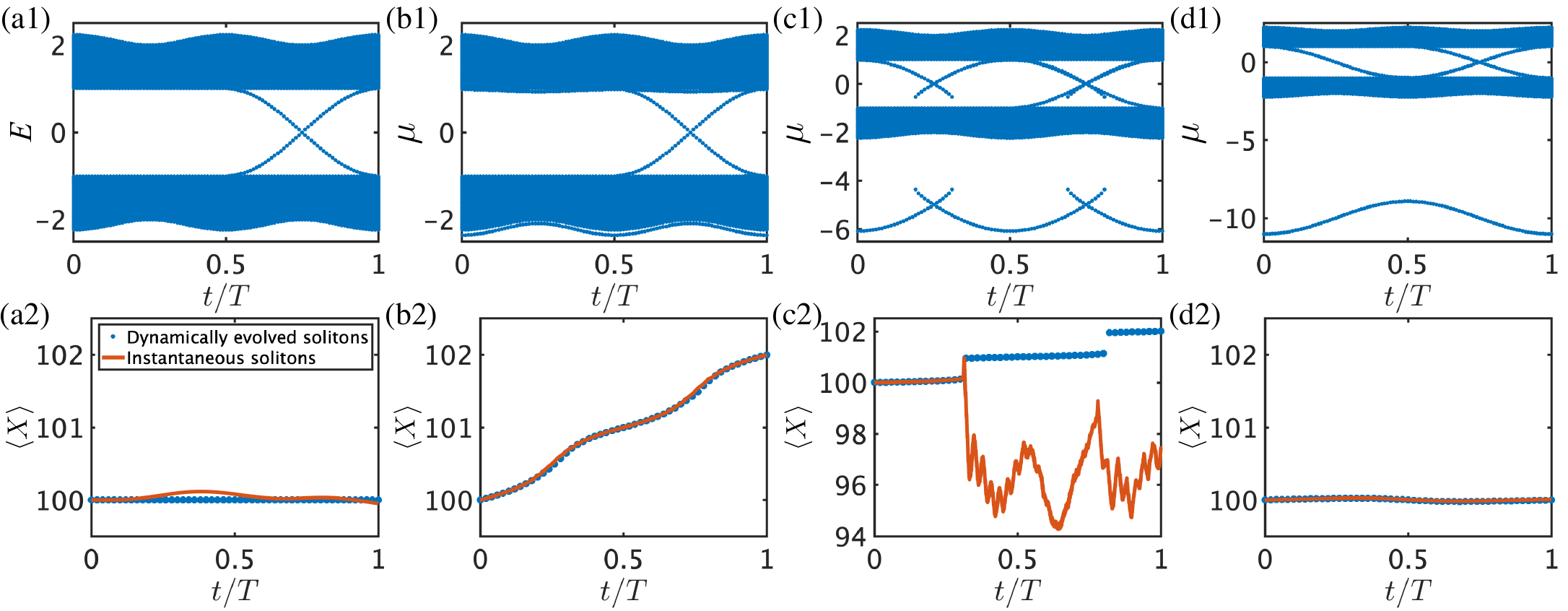}
\caption{\label{fig.1} Nonlinear eigenvalue of nonlinear RM Hamiltonian defined in the second line of Table~\ref{table1} and nonlinear Thouless pumping of soliton. (a1)-(d1) Band structures of the nonlinear eigenvalues for nonlinear RM Hamiltonian (\ref{NRM}). (a2-d2) The expected position of the soliton as a function of time over a single period. The parameters are fixed as $J=1$, $\delta=0.5$, $\Delta=1$, $T=2000\pi$, and $\omega=10^{-3}$. (a1)-(d1)[(a2)-(d2)] The interaction strengths are set to $g=0$, $g=1$, $g=5$, and $g=10$, respectively.}
\end{figure*}

The organizational structure of this paper is outlined as follows. In Sec.~\ref{Section2}, we provide a thorough introduction to the RM Hamiltonian. In Sec.~\ref{Section3}, four different types of eigenvalue problems are defined. In particular, directly inspired by the work of the authors of Ref.~\cite{Isobe2024}, we give the definition of the anomalous eigenvalue’s nonlinearity for the nonlinear RM Hamiltonian and establish an additional form of bulk-edge correspondence based on these auxiliary nonlinear eigenvalues. Section~\ref{Section4} investigate the effects of nonlinear intensity and model parameter on the bulk-edge correspondence of the nonlinear RM model, elucidating how these factors affect the system’s behavior. Finally, in Sec.~\ref{Section5}, we summarize the main results of our investigation.

\section{Nonlinear Rice-Mele Model}
\label{Section2}

In this work, we are interested in a one-dimensional interacting bosonic chain consisting of $N$ dimer units~\cite{Xiao2010}. At the mean-field level, our model system can be well described by the nonlinear RM Hamiltonian\cite{Rice1982, Lin2020, Tuloup2023, Cerjan2020, Kofuji2024, Hayward2018, Walter2023, Kofuji2023, Viebahn2024, Hayward2021, Allen2022, Akira2023, Abhishek2023, Wang2018} as follows:
\begin{eqnarray}
H_{\text{non}}&= & \sum_{n=1}^{2N}\Bigg\{-\Big[J+(-1)^{n}\delta\sin\left(\omega t\right)\Big](\Psi _{n}^{*}\Psi _{n+1}+\text{H.c.})\nonumber\\
 & -&\Delta\cos\left(\omega t\right)\sum_{n=1}^{2N}(-1)^{n}\Psi _{n}^{*}\Psi _{n}-\frac{g}{2}\sum_{n=1}^{2N}\left|\Psi_{n}\right|^{4}\Bigg\}.\label{NRM}
\end{eqnarray}
In Hamiltonian (\ref{NRM}), the $[\Psi_n,\Psi_{n+1}]$ denotes the wave function of the dimer located at site $n$ (two atoms in one unit cell). The $J$ is the uniform hopping amplitude, the $\delta\sin(\omega t)$ is the time-dependent dimerization order, and therefore, the terms $[-J-\delta\sin(\omega t)]$ and $[-J+\delta\sin(\omega t)]$ represent the intracell and intercell coupling, respectively. The $\Delta\cos(\omega t)$ labels the time-dependent staggered sublattice potential, and the $g$ represents the strength of the focusing Kerr-like nonlinearity~\cite{Tuloup2023}. Furthermore, the $\omega$ in Hamiltonian (\ref{NRM}), representing the modulation frequency, is sufficiently small to ensure adiabatic evolution. Finally, we remark that Hamiltonian (\ref{NRM}) is of immediate relevance in the context of recent experiments~\cite{Viebahn2024} for investigating interaction-induced Thouless pumping in a dynamical superlattice.

\begin{table}
\renewcommand\arraystretch{1.7}
\caption{\label{table1}Summary of four kinds of eigenvalues of linear and nonlinear Hamiltonian.}
\begin{ruledtabular}
\begin{tabular}{ccccc}
Equation&Problem definition&Ref.\\ \hline
$\hat{H}_{\text{lin}}\Psi = E\Psi$ &Eigenvalue&\!\!\!\!\!\!\!\!Ref. \cite{Morse1953}\\
$\hat{H}_{\text{non}}\Psi=\mu\Psi$&Nonlinear eigenvalue&\!\!\!\! \!\!\!\!Ref. \cite{Tuloup2023}\\ 
$\hat{H}_{\text{lin}}\Psi=\omega S(\omega)\Psi$&Eigenvalue nonlinearity&\!\!\!\! \!\!\!\!Ref. \cite{Isobe2024}\\
$\!\!\!\!\!\!\hat{H}_{\text{non}}\!\Psi \!\!= \!\omega\! S(\omega) \!\Psi$& \!\!\!\!\!Anomalous eigenvalue nonlinearity& \!\!This work \\
\end{tabular}
\end{ruledtabular}
\end{table}

The purpose and emphasis of this work is to extend the theoretical framework about the eigenvalue’s nonlinearity and the corresponding bulk-edge correspondence of the linear Hamiltonian in Ref.~\cite{Isobe2024} to the counterpart of nonlinear cases, i.e., the anomalous eigenvalue’s nonlinearity and the corresponding anomalous bulk-edge correspondence based on the nonlinear RM Hamiltonian (\ref{NRM}). In this end, we proceed to obtain equations of motion of the nonlinear RM Hamiltonian (\ref{NRM}) by variation of the Hamiltonian (\ref{NRM}) as $i\partial \Psi_j/\partial t=\delta H/\delta \Psi^*_j$ which are described by the following set of nonlinear Schrödinger equations for $j= 0,1,...N-1$,
\begin{eqnarray}
i\frac{\partial\Psi_{2j}}{\partial t}	=& -&\left(J+\delta\sin\omega t\right)\Psi_{2j+1}-\left(J-\delta\sin\omega t\right)\Psi_{2j-1}\nonumber\\
& -&\left[\Delta\cos\omega t+g\left|\Psi_{2j}\right|^{2}\right]\Psi_{2j},\label{GP1}\\
i\frac{\partial\Psi_{2j+1}}{\partial t}=& -&\left(J+\delta\sin\omega t\right)\Psi_{2j}-\left(J-\delta\sin\omega t\right)\Psi_{2j+2}\nonumber\\
& +&\left[\Delta\cos\omega t-g\left|\Psi_{2j+1}\right|^{2}\right]\Psi_{2j+1}.\label{GP2}
\end{eqnarray}
The nonlinear eigenvalue of the nonlinear Hamiltonian defined as $\hat{H}_{\text{non}}\Psi=\mu\Psi$ can be obtained by plugging $\Psi \rightarrow \Psi e^{i\mu t/\hbar}$ into Eqs. (\ref{GP1}) and (\ref{GP2}). We note that Eqs. (\ref{GP1}) and (\ref{GP2}) constitute a class of time-dependent Gross-Pitaevskii equations, or alternatively, discrete nonlinear Schrödinger equations. These equations provide a powerful framework for describing the dynamical behavior of Bose-Einstein condensates within an ultracold quantum gas~\cite{Viebahn2024}, as well as the propagation characteristics of pulsed light through arrays of waveguides~\cite{Tuloup2023}.

\section{Anomalous eigenvalue's nonlinearity of nonlinear Rice-Mele Hamiltonian}
\label{Section3}

\subsection{Four kinds of eigenvalues of linear and nonlinear Hamiltonian}\label{FourEig}

In Sec.~\ref{Section2}, we present the nonlinear RM Hamiltonian (\ref{NRM}) along with the associated discrete nonlinear Schrödinger equations (\ref{GP1}) and (\ref{GP2}). The objective of Sec.~\ref{Section3} is to undertake a thorough examination of the linear and nonlinear eigenvalue problems pertinent to both the linear and nonlinear Hamiltonians of Eq. (\ref{NRM}), drawing direct inspiration from Ref.~\cite{Isobe2024}. Table~\ref{table1} offers a comprehensive overview of the definitions of these problems, serving as a crucial reference for understanding the progression from linear to nonlinear eigenvalue problems within Hamiltonian systems.

We provide an outline of the four distinct types of eigenvalue problems, which are summarized in Table \ref{table1}, as follows.

(i) The eigenvalue of the linear Hamiltonian ($\hat{H}_{\text{lin}}\Psi = E\Psi$) as summarized in the first line of Table \ref{table1}: The nonlinear RM Hamiltonian (\ref{NRM}) simplifies to a linear form when the nonlinearity parameter $g$ is set to zero. In this case, the entire parameter space is spanned by $\delta$ and $\Delta$. The bulk-edge correspondence of the linear RM Hamiltonian exhibits bulk properties along with two edge states, as depicted in Fig.~\ref{fig.1}(a1). The topologically invariant relevant for charge pumping is the Berry phase of the lowest band, which becomes singular at the origin of the $\delta$-$\Delta$ plane \cite{Xiao2010}. Specifically, a trajectory that encloses this singularity corresponds to a nonzero Berry phase and pumps two charges to the neighboring unit cell per pump cycle (equivalent to two atoms in one dimer unit). Conversely, if the trajectory does not enclose the singularity, the Berry phase is zero, and consequently, the pumped charge is also zero.

\begin{figure*}
\includegraphics[width=1\textwidth]{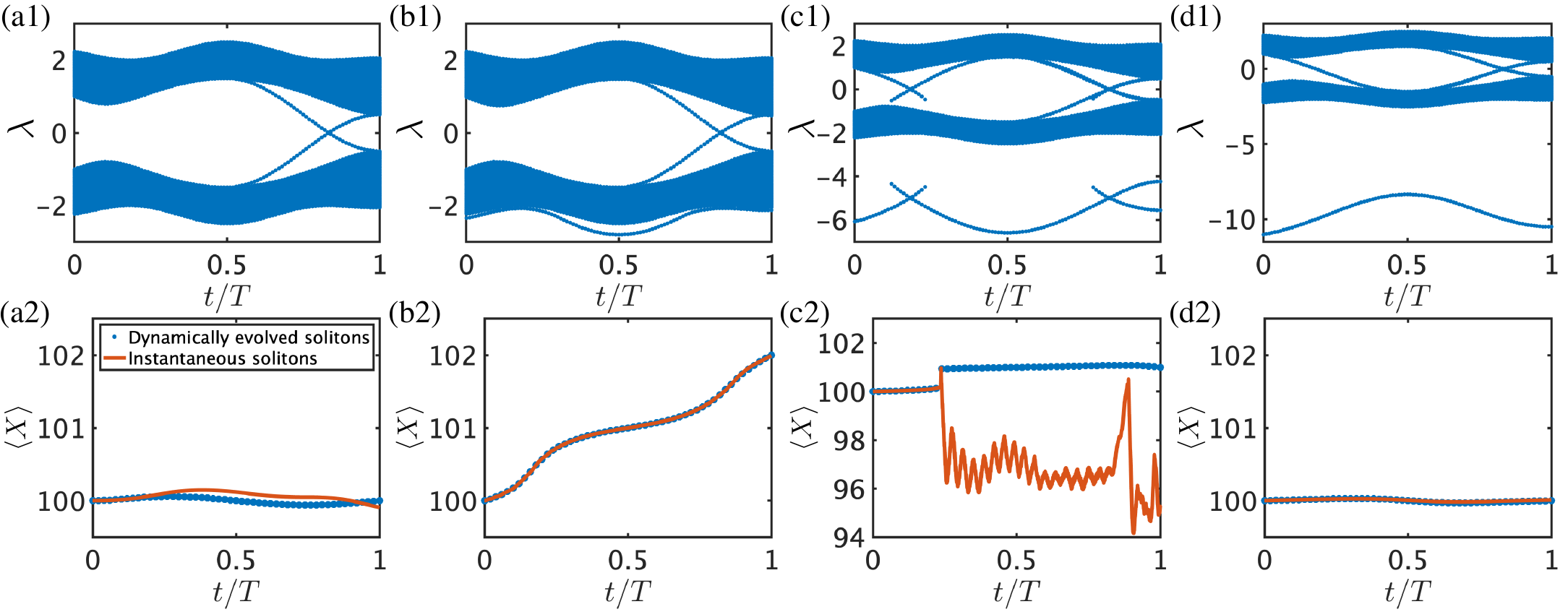}
\caption{\label{fig:2} Eigenvalue's nonlinearity of nonlinear RM Hamiltonian defined the last line of Table \ref{table1} and nonlinear Thouless pumping of soliton. (a1)-(d1): The auxiliary $\lambda$ spectrum for the nonlinear RM Hamiltonian (\ref{NRM}). (a2)-(d2): The anticipated position of the nonlinear excitation of soliton as a function of time over one period. The parameters are fixed at  $J=1$, $\delta=0.5$, $\Delta=1$, $T=2000\pi$ and $\omega=10^{-3}$. Specifically, (a1)-(d1) and their corresponding (a2)-(d2) panels represent interaction strengths of $g=0$, $g=1$, $g=5$, and $g=10$, respectively.}
\end{figure*}

(ii) Nonlinear eigenvalue of the nonlinear Hamiltonian ($\hat{H}_{\text{non}}\Psi=\mu\Psi$) as summarized in the second line of Table \ref{table1}: This eigenvalue problem is derived by substituting $\Psi \rightarrow \Psi e^{i\mu t/\hbar}$ into Eqs. (\ref{GP1}) and (\ref{GP2}). The nonlinear RM Hamiltonian (\ref{NRM}), characterized by a nonzero nonlinearity parameter $g$, gives rise to a plethora of interaction-induced phenomena that have no counterpart in the linear regime. The synergy between topology and interaction has ushered in new avenues for exploring topological transport and bulk-edge correspondence. A quintessential example is the nonlinear Thouless pumping \cite{Jurgensen2021}. In this context, nonlinearity plays a pivotal role in quantizing transport through the formation of solitons and spontaneous symmetry-breaking bifurcations [see Figs. \ref{fig.1}(b2), (c2), and (d2)]. We remark that the discontinuous soliton position jumps in Figs. \ref{fig.1}(c2) and \ref{fig:2}(c2) arise from self-crossing band structures [see Figs. \ref{fig.1}(c1) and \ref{fig:2}(c1)], where adiabaticity breaks due to nonlinearity-induced band's bifurcation at critical nonlinearity. For weak nonlinearity, the motion of the soliton becomes topologically quantized, as illustrated in Figs. \ref{fig.1}(b2). Specifically, the bulk-edge correspondence of the nonlinear RM Hamiltonian (\ref{NRM}) exhibits a quantized displacement by two lattice sites per cycle [see Fig. \ref{fig.1}(b2)], which is directly linked to the topology of the underlying band structures [see Fig. \ref{fig.1}(b1)]. In contrast, no such quantized pumping of solitons occurs when $g=0$, as shown in Fig. \ref{fig.1}(a2). As the nonlinearity strength $g$ increases, self-intersecting bands emerge in the intermediate nonlinear regime with the appearance of the loop structure~\cite{Wu2001} as shown in Fig. \ref{fig.1}(c1), leading to a pumping effect that persists but is no longer quantized, as observed in Fig. \ref{fig.1}(c2). When the nonlinearity further intensifies into the strong nonlinear regime [see Fig. \ref{fig.1}(d1)] and the self-trapped phenomenon is supposed to occur~\cite{Wang2006}, the soliton dynamics are suppressed, resulting in the cessation of the pumping effect, as depicted in Fig. \ref{fig.1}(d2)
 
(iii) The eigenvalue's nonlinearity in a linear Hamiltonian as summarized in the third line of Table \ref{table1}: This scenario introduces a concept where the eigenvalues exhibit a state-variable-dependent nonlinearity.  The eigenvalue equation is given by $\hat{H}_{\text{lin}}\Psi = \omega S(\omega)\Psi$. Here, the Hermitian matrix $H_{\text{lin}}$ and $S$ is the linear Hamiltonian and overlap matrix, respectively,  $\Psi$ is the eigenvector. Note that the matrices $H_{\text{lin}}$ and $S$  may depend on the eigenvalue $\omega$, which is real. Remarkably, the authors of Ref. \cite{Isobe2024} showed that when the nonlinearity is weak but finite, the topological edge states of the auxiliary eigenstates are topologically inherited as physical edge states. This inheritance occurs when the auxiliary eigenvalues are monotonic with respect to the physical eigenvalues. This finding establishes a bulk-edge correspondence in systems with eigenvalue nonlinearity, highlighting the profound implications of such nonlinearity on the topological properties of the system.

(iv) The anomalous eigenvalue's nonlinearity of nonlinear Hamiltonian in the fourth line of Table \ref{table1}: This scenario extends the concept of eigenvalue nonlinearity from linear Hamiltonians in Ref. \cite{Isobe2024}  to their nonlinear counterparts, which is referred to as the anomalous eigenvalue's nonlinearity of the nonlinear Hamiltonian. The definition given by $\hat{H}_{\text{non}}\Psi = \omega S(\omega) \Psi$ represents a generalization where the Hamiltonian is nonlinear. Unlike systems with only wavefunction nonlinearity (e.g., $g|\Psi_n|^2$), the anomalous case involves dual nonlinearity, both the Hamiltonian ($H_{\text{non}}$) and eigenvalues [$\omega S(\omega)$] depend on the state.
This equation models systems with anharmonic potentials, where energy levels and wave functions are intricately intertwined in a nonlinear fashion, providing a richer framework for understanding quantum states.

\subsection{Eigenvalue's nonlinearity and the corresponding bulk-edge correspondence}\label{Eignon}

In Sec.~\ref{FourEig}, we presented an overview of four distinct categories of eigenvalue problems, which are concisely summarized in Table \ref{table1}. In the subsequent Sec.~\ref{Eignon}, we adopt the methodology detailed in Ref.~\cite{Isobe2024} to delve into the anomalous nonlinearity of the eigenvalues within the nonlinear RM Hamiltonian (\ref{NRM}) and its corresponding bulk-edge correspondence. Our strategy for examining this anomalous eigenvalue nonlinearity in the nonlinear RM Hamiltonian (\ref{NRM}) and bulk-edge correspondence, given the underlying nonlinearity of the eigenvalues, involves the utilization of auxiliary eigenvalues, as outlined below.

(i) The anomalous eigenvalue's nonlinearity of nonlinear RM Hamiltonian (\ref{NRM}) is defined as 
\begin{equation}
H_{\text{non}}\left(\omega,t\right)\varPsi=\omega S\left(\omega,t\right)\varPsi.\label{eq:nonlinear}
\end{equation}
In Eq. (\ref{eq:nonlinear}), the time-dependent $H\left(\omega,t\right)$ represents the Hermitian matrix corresponding to the nonlinear RM Hamiltonian (\ref{NRM}). The symbol $\varPsi$ denotes the nonlinear eigenfunction.
Moreover, the $S\left(\omega,t\right)$ is the overlap matrix, and $\omega$ serves as the parameter that characterizes the anomalous nonlinearity compared with the intrinsic nonlinearity of $g$ in the nonlinear RM Hamiltonian (\ref{NRM}). It is worth noting that Eq. (\ref{eq:nonlinear}) together with $S$ are employed to introduce auxiliary eigenvalues and furthermore the topological edge states of auxiliary eigenstates are topologically inherited as physical edge states, which represents interplay between the topology and nonlinearity of eigenvalues.

(ii) Next, we proceed to introduce the matrix $P\left(\omega,\mathbf{k}\right)$ as follows:
\begin{equation}
P\left(\omega,\mathbf{k}\right)=H_{\text{non}}\left(\omega,\mathbf{k}\right)-\omega S\left(\omega,\mathbf{k}\right).\label{EquationP}
\end{equation}
From Eq. (\ref{EquationP}), the solution satisfying the equation $P\left(\omega,\mathbf{k}\right)\varPsi=0$ is equivalent to that of the nonlinear equation presented in Eq. (\ref{eq:nonlinear}). To gain a more profound understanding of the bulk-edge correspondence model encapsulated by this nonlinear equation, we introduce an auxiliary eigenvalue $\lambda$, where $\lambda$ is an element of the real numbers $\mathbb{R}$.
\begin{equation}
P\left(\omega,\mathbf{k}\right)\varPsi = \lambda\varPsi.\label{eq:lambda}
\end{equation}
We remark that  the auxiliary eigenvalue $\lambda$ does not carry physical significance in the general case, with the notable exception when $\lambda=0$. Hence, the core problem reduces to determining the solution of Eq. (\ref{eq:lambda}) specifically at $\lambda=0$.

(iii) Finally, the overlap matrix $S$ appearing in Eq. (\ref{eq:nonlinear}), which exhibits a dependence on the nonlinear parameter $\omega$, is constructed in the following manner:
\begin{eqnarray}
S\left(\omega\right)=\left(\begin{array}{cccc}
S_{0} & 0 & 0 & 0\\
0 & S_{0} & 0 & 0\\
0 & 0 & \ddots & 0\\
0 & 0 & 0 & S_{0}
\end{array}\right),
\end{eqnarray}
with $S_{0}$ being a diagonal matrix given by
\begin{equation}
S_{0}=\left(\begin{array}{cc}
1-M_{S}\left(\omega\right) & 0\\
0 & 1+M_{S}\left(\omega\right)
\end{array}\right),
\end{equation}
and $M_{S}\left(\omega\right)$ is defined as $M_{S}\left(\omega\right)=M_{1}\tanh\left(\omega t\right)/\omega$. It is worth noting that the aforementioned three steps outline the general strategy for deriving the anomalous eigenvalue nonlinearity associated with a nonlinear Hamiltonian, as exemplified in the fourth row of Table \ref{table1}.

We are now in a position to investigate how the interaction, characterized by the parameter $g$ in the Hamiltonian (\ref{NRM}), influences the anomalous eigenvalue nonlinearity. This will be achieved by numerically solving Eqs. (\ref{eq:lambda}). Furthermore, we will demonstrate that the topological edge states of the auxiliary eigenstates are topologically inherited as physical edge states by numerically solving Eqs. (\ref{GP1}) and (\ref{GP2}).

As a preliminary step in solving Eqs. (\ref{eq:lambda}), we fix the parameters $J=1$, $\delta=0.5$, $M_{1}=0.5$. Throughout our calculations, we consider a system comprising $N=100$ unit cells (corresponding to $2N=200$ sites) and initialize the wave function as $\Psi_{0}=\psi_{0}/\sqrt{\int|\psi_{0}|^{2}dx}$ with $\psi_{0}=\cosh^{-1}[\left|x-100\right|/5]$. Using these parameters, we numerically determine the auxiliary $\lambda$ spectrum for various interaction strengths $g$, as depicted in Figs.~\ref{fig:2}(a1) to \ref{fig:2}(d1). Upon introducing nonlinear eigenvalues, we observe that the upper boundary of the lower bulk band and the lower boundary of the upper bulk band no longer remain flat. Instead, they exhibit fluctuations in both upward and downward directions. Notably, the direction of these fluctuations coincides with the direction of the opposite boundary within their respective bulk regions. In contrast to the case with linear eigenvalues, the behaviors of the edge states and soliton states remain largely unchanged, which will be discussed below.

Next, we proceed to investigate the impact of the interaction parameter $g$ on bulk-edge correspondence by studying the evolution of the position expectation value $\langle X\rangle = \sum_{j}j|\Psi_{j}|^2$ for the soliton state over a single adiabatic cycle. To achieve this, we numerically solve Eqs. (\ref{GP1}) and (\ref{GP2}), employing two distinct methodologies. In the first approach, we numerically propagate the initial soliton profile through time evolution governed by the nonlinear Schrödinger equations [Eqs. (\ref{GP1}) and (\ref{GP2})], implemented via a fourth-order Runge-Kutta algorithm. This method directly simulates the dynamical evolution of the wave function under physical time-dependent driving. In the second approach, we employ an iterative self-consistent scheme to converge to the steady-state solution at each time slice $t$, bypassing explicit time propagation. Under adiabatic conditions ($\omega\rightarrow 0$), both methods yield identical results, thereby mutually validating their accuracy. For nonadiabatic driving ($\omega \gg 0$), the first method faithfully reproduces experimentally observable dynamics [e.g., soliton jumps in Figs. \ref{fig.1}(c2) and \ref{fig:2}(c2)], while the second method serves as a diagnostic tool to quantify deviations from adiabaticity by comparing instantaneous eigenstates with dynamically evolved states. We remark that in the second approach, the wave function evolves through iterative steps as an iterative term, thereby ensuring the successful derivation of the band structure for the nonlinear wave function. It is important to highlight that when the eigenvalues of the RM model transition to nonlinear eigenvalues, the pumping behavior of the system exhibits significant changes. Specifically, at $g=0$, the pump transport oscillates near the initial point, as shown in Fig.~\ref{fig:2}(a2). As the interaction strength increases to $g=1$ and $g=5$, the system demonstrates pump transport phenomena. When $g=1$, the pump transport value over one period remains 2, as depicted in Fig.~\ref{fig:2}(b2). Furthermore, when the interaction strength continues to rise to $g=5$, the soliton wave dynamical evolve within a period decreases more pronouncedly, as depicted in Fig.~\ref{fig:2}(c2). Consistent with the original RM model, when $g=10$, the pumping ceases to operate in the RM model with the nonlinear eigenvalue problem, as illustrated in Fig.~\ref{fig:2}(d2).

Finally, note that this section provides a thorough examination of the eigenvalue problem in Hamiltonian systems, contrasting the linear and nonlinear cases. Linear Hamiltonians serve as the cornerstone for determining energy states in both classical and quantum mechanics. In contrast, nonlinear Hamiltonians, which depend on the magnitude of the wave function, introduce more intricate challenges. We also discuss how the incorporation of eigenvalue terms that depend on state variables enriches linear Hamiltonian systems, giving rise to phenomena such as level crossing. Furthermore, we extend the concept of nonlinear eigenvalues to nonlinear Hamiltonian systems, emphasizing their vital role in modeling complex quantum states described by nonharmonic potentials. By introducing an auxiliary eigenvalue $\lambda$, we derive the time-energy spectrum of $\lambda$ and investigate the temporal evolution of the expectation value of the position for the ground state. Our findings reveal that the inclusion of nonlinear eigenvalues primarily influences the pump transmission and stability of the ground-state solitons.

\section{Nonlinear bulk-edge correspondence}
\label{Section4}

\begin{figure}
\includegraphics[width=1\columnwidth]{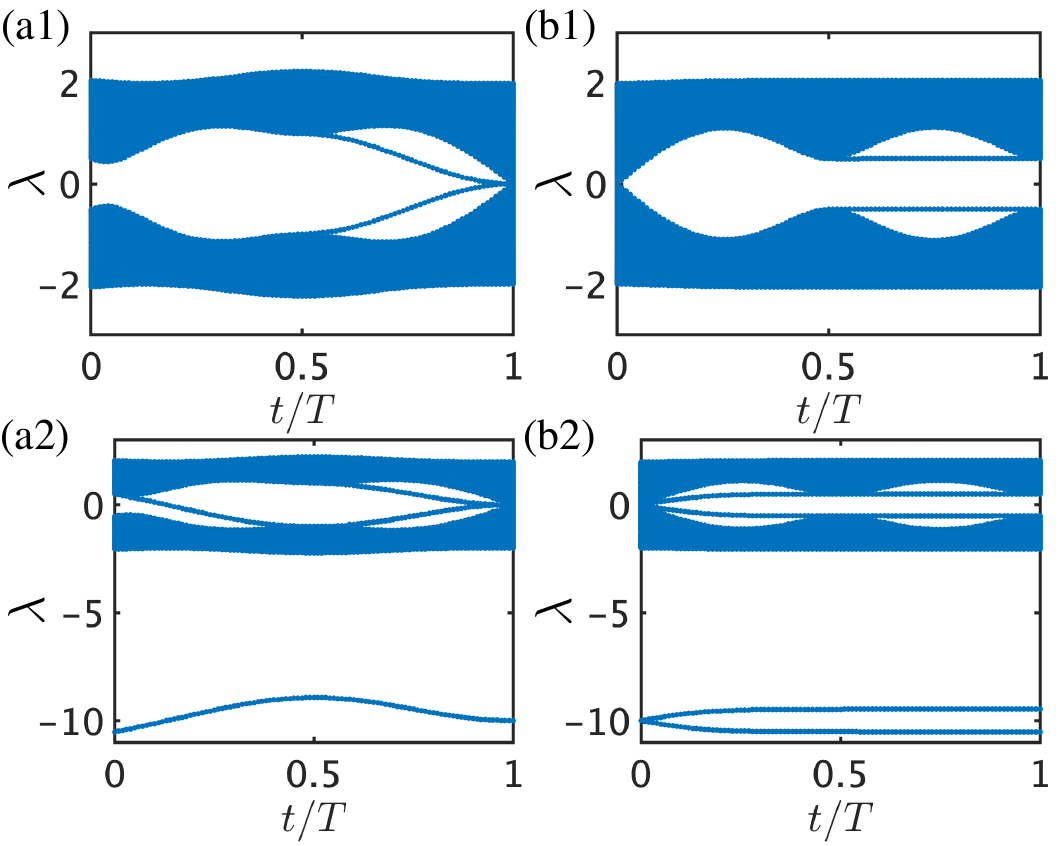}
\caption{\label{fig:3} Effects of parameter $\Delta$ and $g$ on the auxiliary $\lambda$ spectrum defined in Eq. (\ref{eq:lambda}) for the nonlinear RM Hamiltonian (\ref{NRM}).The parameters are fixed
at $J=1$, $\delta=0.5$, $T=2000\pi$, and $\omega=10^{-3}$. Other parameters are given as (a1): $g=0$, $\Delta=0.5$; (a2): $g=0$, $\Delta=0$; (b1): $g=10$, $\Delta=0.5$; (b2): $g=10$, $\Delta=0$.}
\end{figure}

In the preceding Sec.~\ref{Section3}, we examined four distinct eigenvalue problems and computed the anomalous nonlinear eigenvalue problem for the nonlinear RM Hamiltonian (\ref{NRM}), obtaining its $\lambda$ spectrum and the pumping diagram for the ground state. Following this, Sec.~\ref{Section4} presents a comprehensive analysis of how the parameter $\Delta$ influences the anomalous nonlinear eigenvalue problem in the nonlinear RM Hamiltonian (\ref{NRM}).

To gain a deeper understanding of the anomalous nonlinear eigenvalues in the nonlinear RM Hamiltonian (\ref{NRM}), we investigate the influence of two key parameters: the interaction strength $g$ and the model parameter $\Delta$, on its energy spectrum. We begin by setting the interaction strength $g$ to 0, allowing us to observe the effect of varying the model parameter $\Delta$ on the eigenvalue of the nonlinearity energy spectrum.
Next, we introduce the interaction strength $g$. Given that the adiabatic process remains valid only in the regimes of weak and strong interactions, this paper focuses exclusively on these two cases. However, since the graphical representation for weak interactions is less pronounced, we opt to explore the case of strong interaction, setting $g=10$ as an illustrative example.

In the first scenario, we consider the case where the interaction strength $g$ is set to 0. As the model parameter $\Delta$ is decreased from an initial value of 1 to 0.5, the intersection points of the edge states gradually shift to the right, ultimately aligning with the end of a period, as illustrated in Fig.~\ref{fig:3}(a1). It is worth noting that these intersection points correspond to $\lambda=0$. As $\Delta$ is further reduced from 0.5 towards 0, the edge states tend to align in parallel, and the points where $\lambda=0$ diminish. Concurrently, the band gap near the initial time position progressively narrows, eventually closing completely at $\Delta=0$. At this point, a new $\lambda=0$ point emerges, as depicted in Fig.~\ref{fig:3}(b1).

In the second scenario, we examine the case where the interaction strength $g$ is set to 10. As the model parameter $\Delta$ is decreased from an initial value of 1 to 0.5, the intersection points of the edge states shift towards the end of the period, mirroring the behavior observed when $g=0$, as illustrated in Fig.~\ref{fig:3}(a2). As $\Delta$ is further reduced from 0.5 towards 0, the edge states in the latter half of the period begin to diverge, and the $\lambda=0$ point becomes a singular point. At $\Delta=0.3285$, an additional edge state emerges in the first half of the period, accompanied by the appearance of a new soliton state in the vicinity of the ground state. Upon further reducing $\Delta$ to 0, we obtain the band structure shown in Fig.~\ref{fig:3}(b2).

\begin{figure}

\includegraphics[width=1\columnwidth]{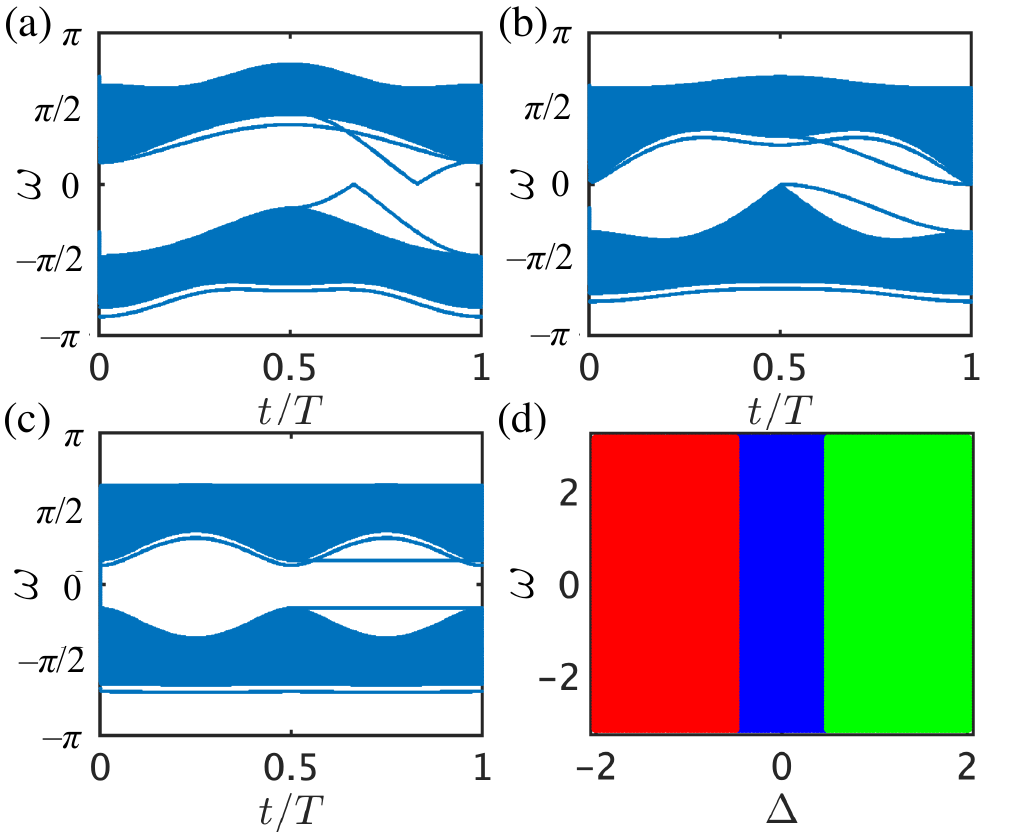}
\caption{\label{fig:4}(a)-(c) Characterization of the relationship between time
$t$ and frequency $\omega$ at $\lambda=0$ for the RM model
with $g=2$. Parameter $\Delta$ are set to $\Delta=1$,
$\Delta=0.5$, and $\Delta=0$, respectively. (d) The Chern numbers
calculated for each $\Delta$ and $\omega$, red indicates a Chern
number $\mathcal{C}=1$; blue indicates $\mathcal{C}=0$; green indicates
$\mathcal{C}=-1$.}

\end{figure}

To investigate the influence of the model parameter $\Delta$ on the nonlinearity of the anomalous eigenvalue associated with the nonlinear RM Hamiltonian [Eq. (\ref{NRM})], with a specific focus on its effect on the $t-\omega$ relationship, we will generate $t-\omega$ diagrams for a range of $\Delta$ values while maintaining $\lambda=0$. Meanwhile, it is worth highlighting that soliton states emerge independently of the specific value of $\Delta$. Moreover, the count of soliton states escalates with the interaction strength g. To facilitate a clearer understanding, we focus our examination on the scenario where $g=2$, which results in the presence of two soliton states, thereby enabling a more straightforward analysis and observation of our findings.

 We proceed to plot the $t-\omega$ diagrams for different values of $\Delta$ ($\Delta=1$, $\Delta=0.5$, and $\Delta=0$), as shown in Figs.~\ref{fig:4}(a) to \ref{fig:4}(c) under the conditions of $\lambda=0$ and $g=2$. When $\Delta=1$, the edge states exhibit a pronounced turning point, as illustrated in Fig. \ref{fig:4}(a), which is in accordance with the edge state crossing observed in Fig. \ref{fig:2}(a1). It is observed that in the band diagram of Fig. \ref{fig:4}(a), the edge states exhibit sharp bending due to the emergence of self-crossing structures within the energy bands of the nonlinear Bloch system. In contrast, for $\Delta=0.5$, the edge states cease to display any turning behavior, as depicted in Fig. \ref{fig:4}(b). This observation aligns with the convergence of edge states at the conclusion of the time period shown in Figs.~\ref{fig:3}(a1) and \ref{fig:3}(a2), hinting at the occurrence of a nonlinear topological phase transition. Lastly, when $\Delta=0$, the edge states remain uncrossed and retain a parallel configuration, as demonstrated in Fig. \ref{fig:4}(c). This result is consistent with the parallel edge states observed in Figs. \ref{fig:3}(b1) and \ref{fig:3}(b2).
 
To elucidate the observed variations in the edge and soliton states presented in the aforementioned figures, we employ the topological invariant of the system, namely the Chern number associated with the RM model. The Chern number for the first nondegenerate band of a two-dimensional system is mathematically defined as the integral of the Berry curvature over the entire Brillouin zone, which can be expressed as follows~\cite{Xiao2010,Isobe2024}:
\begin{equation}
\mathcal{C}_{1}=\frac{1}{2\pi}\int_{0}^{2\pi}dk\int_{0}^{T}dt \left[\nabla_{\vec{R}}\times\mathcal{A}_{1}(\vec{R})\right],
\end{equation}
where $\mathcal{A}_{1}(\vec{R})=i\langle \Psi_{1}(k,t) | \nabla_{k} | \Psi_{1}(k,t)\rangle $ is the Berry connection for the ground state. Here, $| \Psi_{1}(k,t) \rangle$ is the wave function in the momentum space of $k$ 
by making a Fourier transformation of the wave function $|\psi_n(t)\rangle$ in Eq. (\ref{NRM}).  We remark that for a one-dimensional dimerized lattice (with two sites per unit cell), the transformation of the discrete real-space wave function 
$ | \Psi_n(t)\rangle$ into momentum $k$ space requires careful treatment of the unit cell structure. Consider a system with $N$ unit cells, where each unit cell contains two lattice sites (indexed as 
$n=1,2,…,2N$). We define the unit cell index $j=1,2,..., N$ and partition the lattice into even-indexed and odd-indexed site sublattices. The momentum-space wave functions for each sublattice are obtained via a discrete Fourier transformation: $|\Psi_{A}(k,t)\rangle= 1/\sqrt{N}\sum_{j=1}^Ne^{ikj}|\Psi_{2j}(t)\rangle$ and $| \Psi_{B}(k,t)\rangle= 1/\sqrt{N}\sum_{j=1}^Ne^{ikj}|\Psi_{2j-1}(t)\rangle$ with $k =2\pi m/N$  ($m=0,1,...,N-1$). The total momentum-space wave function is then expressed as a two-component spinor as $ | \Psi_{1}(k,t)\rangle=(| \Psi_{A}(k,t), | \Psi_{B}(k,t))^T$. Here since the model system is one-dimensional, $k$ is a scalar. The vector $\vec{R}$  represents the integration boundaries, specifically $\vec{R}=\left[0;2\pi\right]\times\left[0;T\right]$.
In the RM model at $g=0$, with the parameter $\omega$ fixed at $10^{-3}$, it is observed that the Chern number adopts nonzero values when $\left|\Delta\right|>0.5$. Our numerical computations reveal that the ground-state band presented in Fig.~\ref{fig:2}(a1) exhibits a Chern number of $\mathcal{C}_{1}=1$. In contrast, the ground state as shown in Fig.~\ref{fig:3}(b1) shows a Chern number of $\mathcal{C}_{1}=0$. In particular, the ground state as showed in Fig.~\ref{fig:3}(a1) shows the critical points of Chern number. These results strongly indicate a profound correlation between the value of the Chern number and the count of points where $\lambda=0$.

To delve deeper into the connection between the Chern number and the band structure of the anomalous eigenvalue's nonlinearity of nonlinear RM model, we construct a phase diagram in the $\Delta -\omega$ parameter space, as shown in Fig.~\ref{fig:4}(d). This phase diagram encapsulates two distinct phases of the nonlinear RM model this diagram illustrates a topologically nontrivial phase, identified by a nonzero Chern number $\mathcal{C}_{1}$ (shown in red and green regions), coexisting with a topologically trivial phase where $\mathcal{C}_{1}=0$ (blue regions). Specifically, the green region corresponds to $\mathcal{C}_{1}=-1$.
By fixing $\omega =10^{-3}$ and drawing the corresponding line on the phase diagram, we can identify the intersection points with the phase boundaries as the locations where changes occur in the edge states. These observations underscore a profound connection between the band structure of the nonlinear RM model and the Chern number.

Therefore, it is concluded that the eigenvalue's nonlinearity of nonlinear RM model exhibits a remarkable capability to modulate the behavior of edge states within the time-energy spectrum through the adjustment of the parameter $\Delta$. Specifically, when $g=0$, a decrease in $\Delta$ from 1 to 0.5 induces a rightward shift in the intersections of edge states during the latter portion of the period. As $\Delta$ continues to diminish towards 0, these intersections gradually diminish, ultimately culminating in the closure of the band gap at $t=0$ when $\Delta=0$.
Similarly, for $g=10$, analogous behavior is observed. However, a distinctive phenomenon emerges at $\Delta=0.3285$, characterized by the appearance of additional bands. This manifestation is evidenced by the emergence of an extra soliton state in proximity to the ground state and the appearance of edge states during the first half of the period, both of which are intricately connected to the Chern number. When $g=0$ and $|\Delta|>0.5$, the Chern number adopts nonzero values, and the model undergoes a nonlinear topological phase transition precisely at $\Delta=0.5$.
The phase diagram in the $\Delta-\omega$ parameter space delineates the boundaries separating topologically nontrivial and trivial phases. The intersections of these phases at specific $\omega$ values unveil alterations in the edge states, further underscoring the intricate interplay between the Chern number and the band structure of the nonlinear RM model.

\section{Conclusion and outlook}
\label{Section5}

This work delves into the behavior of the RM model across varying degrees of nonlinearity, unveiling the intricate interplay between topological attributes and nonlinear dynamical processes. The research unequivocally demonstrates a transition from a linear Hamiltonian regime, typified by a topological insulator harboring protected edge states, to a regime manifesting solitons and nonlinear pumping phenomena upon the introduction of nonlinearity. By incorporating an auxiliary eigenvalue $\lambda$, the study offers a vantage point on the time-energy spectrum and the temporal progression of the ground state, underscoring the pivotal role of the nonlinear eigenvalue in governing pumping transport and stability.

The study elucidates the anomalous conduct of the nonlinear RM model, showcasing its capacity to modulate edge states within the time-energy spectrum through adjustments to the parameter $\Delta$. Notably, the model undergoes a nonlinear topological phase transition at $\Delta=0.5$, with the Chern number serving as a crucial indicator of topological metamorphoses. The phase diagram in the $\Delta-\omega$ parameter space distinctly demarcates topologically nontrivial and trivial phases, and the intersections of these phases at specific $\omega$ values provide invaluable insights into the alterations in edge states.
This work constitutes a substantial contribution to the comprehension of the RM model's behavior under nonlinear conditions, revealing the myriad phenomena that emerge from the synergy of nonlinearity and topology. The findings carry profound implications for the design of topological insulators and the manipulation of edge states in quantum systems, opening up promising avenues for future research endeavors and technological advancements.

\section*{Acknowledgements}
\label{Section6}

We thank Ying Hu, Yapeng Zhang, Nan Li, Shujie Cheng, Chenhui Yan, and Biao Wu for stimulating discussions and useful help. This research was supported by the Zhejiang Provincial Natural Science Foundation of China under Grant No. LZ25A040004 and the National Natural Science Foundation of China under Grants No. 12074344. 

\section*{Data availability}
\label{Section7}
	
The data that support the findings of this article are not publicly available upon publication because it is not technically feasible and/or the cost of preparing, depositing, and hosting the data would be prohibitive within the terms of this research project. The data are available from the authors upon reasonable request.
	
\bibliography{Reference}

\end{document}